%% file: paper.tex
\PassOptionsToPackage{dvipsnames}{xcolor}
\documentclass[sigconf,authorversion,screen]{acmart}

\usepackage[all]{nowidow}
\usepackage{xcolor}
\usepackage{bm}
\usepackage{subcaption}
\usepackage{hyperref}
\usepackage{acronym}
\usepackage[capitalise,noabbrev]{cleveref}

\crefformat{section}{\S#2#1#3}
\crefformat{subsection}{\S#2#1#3}
\crefformat{subsubsection}{\S#2#1#3}
\crefrangeformat{section}{\S#3#1#4 to~\S#5#2#6}
\crefrangeformat{subsection}{\S#3#1#4 to~\S#5#2#6}
\crefrangeformat{subsubsection}{\S#3#1#4 to~\S#5#2#6}

\newcommand{\sysname}{\textrm{LLM4FaaS}}
\copyrightyear{2025}
\acmYear{2025}
\setcopyright{cc}
\setcctype{by-nc}
\acmConference[UCC '25]{2025 IEEE/ACM 18th International Conference on Utility and Cloud Computing}{December 1--4, 2025}{Nantes, France}
\acmBooktitle{2025 IEEE/ACM 18th International Conference on Utility and Cloud Computing (UCC '25), December 1--4, 2025, Nantes, France}\acmDOI{10.1145/3773274.3774686}
\acmISBN{979-8-4007-2285-1/2025/12}

\begin{document}

\author{Minghe Wang}
\affiliation{%
    \institution{TU Berlin}
    \city{Berlin}
    \country{Germany}}
\email{mw@3s.tu-berlin.de}
\orcid{0009-0001-3780-5828}

\author{Tobias Pfandzelter}
\affiliation{%
    \institution{TU Berlin}
    \city{Berlin}
    \country{Germany}}
\email{tp@3s.tu-berlin.de}
\orcid{0000-0002-7868-8613}

\author{Trever Schirmer}
\affiliation{%
    \institution{TU Berlin}
    \city{Berlin}
    \country{Germany}}
\email{ts@3s.tu-berlin.de}
\orcid{0000-0001-9277-3032}

\author{David Bermbach}
\affiliation{%
    \institution{TU Berlin}
    \city{Berlin}
    \country{Germany}}
\email{db@3s.tu-berlin.de}
\orcid{0000-0002-7524-3256}

\title{LLM4FaaS: No-Code Application Development using LLMs and FaaS}
\keywords{Large Language Models, Function-as-a-Service, No-Code Development}

\begin{abstract}
    Large language models (LLMs) show great capabilities in generating code from natural language descriptions, bringing programming power closer to non-technical users.
    However, their lack of expertise in operating the generated code remains a key barrier to realizing customized applications.
    Function-as-a-Service (FaaS) platforms offer a high level of abstraction for code execution and deployment, allowing users to run LLM-generated code without requiring technical expertise or incurring operational overhead.

    In this paper, we present \sysname{}, a no-code application development approach that integrates LLMs and FaaS platforms to enable non-technical users to build and run customized applications using only natural language.
    By deploying LLM-generated code through FaaS, \sysname{} abstracts away infrastructure management and boilerplate code generation.  
    We implement a proof-of-concept prototype based on an open-source FaaS platform, and evaluate it using real prompts from non-technical users.
    Experiments with \emph{GPT-4o} show that \sysname{} can automatically build and deploy code in 71.47\% of cases, outperforming a non-FaaS baseline at 43.48\% and an existing LLM-based platform at 14.55\%, narrowing the gap to human performance at 88.99\%.
    Further analysis of code quality, programming language diversity, latency, and consistency demonstrates a balanced performance in terms of efficiency, maintainability and availability.
\end{abstract}

\ccsdesc[500]{Computing methodologies~Natural language processing}
\ccsdesc[500]{Computer systems organization~Cloud computing}
\ccsdesc[500]{Human-centered computing~User centered design}

\maketitle

\input{sections/1_introduction.tex}
\input{sections/2_background.tex}
\input{sections/3_architecture.tex}

\input{sections/4_evaluation.tex}
\input{sections/6_conclusion.tex}

\begin{acks}
Partially funded by the \grantsponsor{BMFTR}{Bundesministerium für Forschung, Technologie und Raumfahrt (BMFTR, German Federal Ministry of Research, Technology and Space)}{https://www.bmftr.bund.de/EN/Home/home\_node.html} -- \grantnum{BMBF}{16KISK183} and \grantnum{BMFTR}{01IS23068}.
We thank Aris Wiegand for their thoughtful comments on this paper.

\end{acks}

\bibliographystyle{ACM-Reference-Format}
\bibliography{bibliography-short.bib}

\balance
\end{document}

%% file: sections/1_introduction.tex
\section{Introduction}
\label{sec:introduction}
Large language models (LLMs) have shown remarkable capabilities in processing natural language requests and generating corresponding code, thus bridging the gap between non-technical users and the technical world~\cite{liu2024your,vaithilingam2022expectation,ni2023lever,weisz2021perfection,xu2022ide,jin2024can,smith2020unleashing}.
However, while non-technical users can use LLMs to generate code for their desired functionality, they typically lack the expertise to properly deploy and run the generated code.
For most people, managing servers, configuring services, or even using the command line are high barriers to operating applications.
We believe that the Function-as-a-Service (FaaS) paradigm and its no-ops principle can help:
FaaS platforms offer a scalable, event-driven, and fine-grained infrastructure abstraction~\cite{gupta2023integration,macia2023serverless,paper_bermbach2021_cloud_engineering,kjorveziroski2021iot,gadepalli2019challenges,wen2021empirical,wolski2019cspot,pfandzelter2023serverless,schirmer2023nightshift,wang2023lotus,malekabbasi2024geofaas}.
By decoupling functionality from infrastructure management, FaaS aligns with the principle of separation of concerns in application development, allowing developers to focus on business logic rather than operational concerns.

In this paper, we propose combining the capabilities of LLMs with the abstractions provided by FaaS to enable non-technical users to build and operate custom applications solely through natural-language descriptions.
%
For this, we present \sysname{}, a no-code application development approach for non-technical end-users, leveraging (i)~the natural language processing capabilities of LLMs to transform user requirements into code snippets and (ii)~FaaS abstractions to streamline code generation, accelerating and simplifying the application development process.
In this way, \sysname{} enables both application customization and development efficiency.
With a 71.47\% semantic and 87.55\% syntax pass rate, \sysname{} demonstrates promising feasibility and consistent performance in our evaluation.
The average end-to-end latency of \sysname{}, from LLM generation to successful function deployment on the FaaS platform, is 23.18 s, with LLM generation contributing the most, i.e., 15.53 s, highlighting the streamlined and efficient nature of \sysname{}.

Overall, we make the following contributions:

\begin{itemize}
    \item We introduce \sysname{}, a novel no-code application development approach based on LLMs and FaaS (\S\ref{sec:architecture}).
    \item We implement a proof-of-concept prototype using  \emph{tinyFaaS}~\cite{paper_pfandzelter2020_tinyfaas}, an open-source, lightweight FaaS platform.  (\S\ref{sec:eva:poc}).
    \item We evaluate the performance of \sysname{} with \emph{GPT-4o}~\cite{hurst2024gpt} and compare with (i)~LLM-generated code running outside FaaS, (ii)~an LLM-based code generation and execution platform, and (iii)~human developers, focusing on pass rates, latency, and code quality (\S\ref{sec:eva:results}).
    \item We provide further insights of performance across programming languages and execution consistency.
\end{itemize}

We make all artifacts and dataset used to produce this paper available publicly. 

%% file: sections/2_background.tex
\section{Background and Related Work}
\label{sec:background}
\sysname{} builds upon two main components: LLMs for natural language based code generation and FaaS platform for application deployment.
\subsection{LLMs in Software Engineering}
\label{sec:background:llms}
LLMs are deep learning models trained on extensive text corpora to understand, generate, and manipulate human language. 
LLMs excel in a wide range of natural language tasks, including text generation, translation, and code-related activities, e.g., generation, modification, and verification, making them valuable tools for software engineering~\cite{liu2024your,vaithilingam2022expectation,ni2023lever,weisz2021perfection,xu2022ide,jin2024can,bernsteiner2022citizen}. 
Although the advanced natural language interpretation ability of LLMs is a promising avenue for the involvement of individuals without programming skills in software development~\cite{bernsteiner2022citizen,smith2020unleashing,corradini2021floware}, operational concerns remain a considerable barrier.

\subsection{Function-as-a-Service (FaaS)}
\label{sec:background:faas}
FaaS is a cloud computing model that offers a flexible and fine-grained abstraction of infrastructure~\cite{jonas2019cloud,baldini2017serverless,mcgrath2017serverless}, thereby minimizing operational overhead and enabling developers to focus on function logic. 
With capabilities, e.g., scale-to-zero, event-driven execution, and on-demand scaling, FaaS is particularly well-suited for IoT and other dynamic application domains~\cite{gupta2023integration,macia2023serverless,kjorveziroski2021iot,gadepalli2019challenges,wen2021empirical,wolski2019cspot}. 
These features support real-time responsiveness to changing workloads, accelerate development, reduce time-to-market, and facilitate rapid iteration on key functionality.

\subsection{Related Work}
\label{sec:rw}
Prior work has integrated LLMs into low-code environments to support human-AI collaboration. 
For example, Liu et al.~\cite{liu2024empirical} empirically compare LLM-based and traditional low-code development by analyzing Stack Overflow discussions, showing that LLM assistance enables a broader range of automation scenarios.
Cai et al.~\cite{cai2023low} introduce Low-Code LLM, which combines planning and executing LLMs with a graphical workflow interface to enhance human-LLM interaction in complex tasks.
Buchmann et al.~\cite{buchmann2024white} and Hagel et al.~\cite{hagel2024turning} similarly leverage LLMs in model- and DSL-based low-code settings. 
While these works show potential for accelerating code refinement and model generation, they require prior knowledge or rely on technical supervision and refinement, limiting accessibility for non-technical participants.

Other efforts move toward no-code paradigms, where LLMs directly generate executable structures from natural language. 
Gao et al.~\cite{gao2024chatiot} propose a zero-code approach for generating trigger-action programs in smart home automation. 
Esashi et al.~\cite{esashi2024action} use LLMs to generate FaaS workflows, assisting cloud developers in configuring serverless applications.
Khojah et al.~\cite{khojah2025impact} investigate prompt engineering strategies that affect the correctness of LLM-generated functions.
Eskandani et al.~\cite{eskandani2024towards} discuss the role of LLMs in addressing FaaS challenges such as cold starts and statelessness, and Kathiriya et al.~\cite{kathiriyaserverless} outline a conceptual LLM-enabled architecture for banking applications. 
While these studies show the feasibility of LLMs-based no-code development, most of them remain intermediate artifacts generation and depend on infrastructure knowledge. 

Beyond academic prototypes, practical notebook-based systems, e.g., Google Colab, Replit, also demonstrate the feasibility of interactive code generation and execution.
These systems enable instant code execution within managed sessions, targeting ephemeral experimentation and requiring manual control, whereas our work focuses on automated, persistent, and event-driven deployment workflows.

To the best of our knowledge, \sysname{} is the first no-code application development system that unifies LLMs with FaaS, enabling non-technical users to build functional applications directly from natural language descriptions.

%% file: sections/3_architecture.tex
\section{Architecture}
\label{sec:architecture}
\newcommand{\bridge}{\textrm{LLM-FaaS Bridge}}

\begin{figure}
    \centering
    \includegraphics[width=0.98\linewidth]{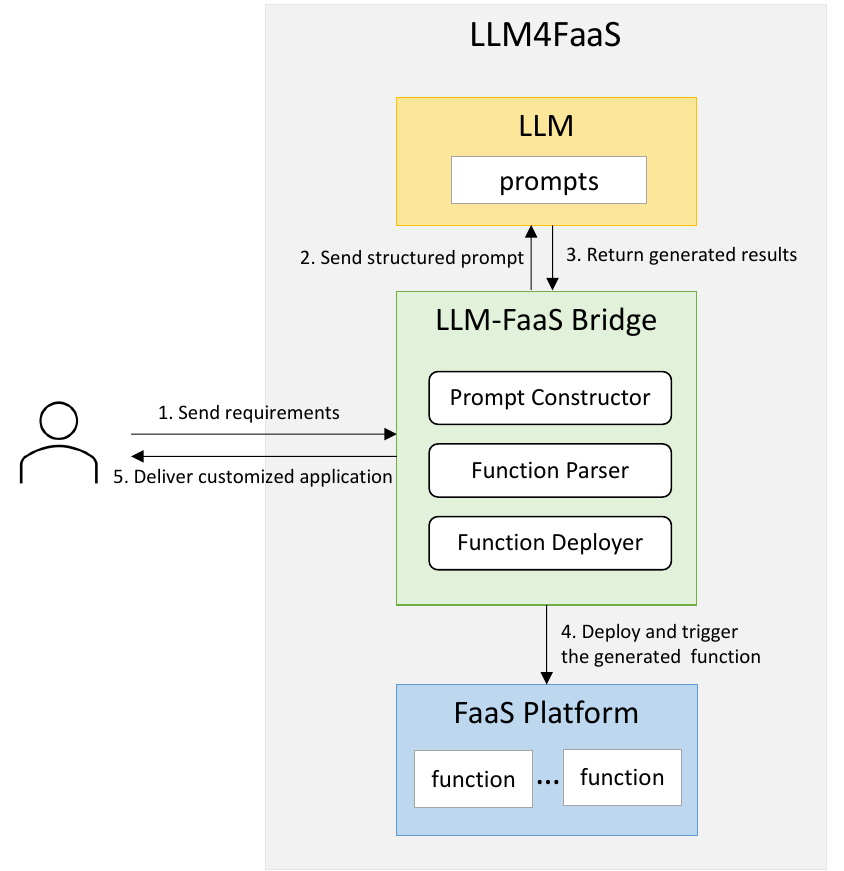}
    \caption{
        \sysname{} consists of three components: an LLM system, a FaaS system, and \bridge{}, the orchestration system. 
        Together, they integrate the power of LLMs and the abstraction of FaaS to enable end-to-end application customization for non-technical users.
    }
    \label{fig:arch}
\end{figure}

With \sysname{}, we aim to empower non-technical users to have customized applications by only providing natural language descriptions.
We show the system design of \sysname{} in Figure~\ref{fig:arch}.

\sysname{} consists of three interconnected systems:
(i) an LLM system that interprets user intents and generates code from structured prompts, 
(ii) a FaaS system that supports scalable and event-driven function deployment, and 
(iii) an orchestration system, \bridge{}, which coordinates the interaction between the two and also serves as the interface for user interaction.
Both the LLM and FaaS systems can be hosted via cloud services or deployed locally.
This modular architecture clearly separates concerns between code generation, deployment, and coordination. 
By integrating with a FaaS platform, \sysname{} eliminates the need for infrastructure and operation concern, enabling a streamlined and fully automated application lifecycle.

\bridge{} is the core system of \sysname{}, consisting of three key components, i.e., \emph{Prompt Constructor}, \emph{Function Parser}, and \emph{Function Deployer}.
Specifically, when a user provides a functional description to \sysname{}, the \bridge{} receives and assigns it to \emph{Prompt Constructor} (Step 1).
\emph{Prompt Constructor} transforms the user description into a structured prompt by incorporating contextual details, e.g., API references, runtime environment, and application constraints. 
This ensures that the LLM receives sufficient information to generate accurate and functional code.
\emph{Prompt Constructor} passes the structured prompt to LLM, waiting for response, which contains the corresponding function logic (Step 2). 
Then, \emph{Function Parser} fetches the LLM-generated output, extracts the function code, and prepares auxiliary FaaS deployment assets, e.g., packaging required dependencies and incorporating application source files (Step 3). 
The processed function is then handed off to the \emph{Function Deployer}, which interacts with the FaaS platform for function registration and deployment (Step 4).
Finally, once the function is successfully deployed, it becomes available for invocation (Step 5). 
Users can interact with the deployed function through predefined interfaces, while the FaaS platform manages execution, scaling, and resource allocation. 
In this way, \sysname{} enables seamless transformation from high-level user intent to executable, fully-managed, and customized applications.

%% file: sections/4_evaluation.tex
\section{Evaluation}
\label{sec:eva}
We evaluate \sysname{} through two complementary approaches.
First, we demonstrate the feasibility of \sysname{} via a proof-of-concept prototype (\S\ref{sec:eva:poc}).
Second, we assess the extent to which our approach can generate ready-to-use applications from natural language descriptions.
To this end, we collect a dataset of natural language application descriptions from a group of non-technical users, which we use on our prototype to evaluate the efficacy of \sysname{} (\S\ref{sec:eva:study-design}).
We present the evaluation results in \S\ref{sec:eva:results} and discuss our findings in \S\ref{sec:eva:discussion}.
    
\begin{figure*}
    \centering
    \subfloat[Syntactic pass rate\label{fig:eva:syntactic}]{
        \includegraphics[width=0.49\linewidth]{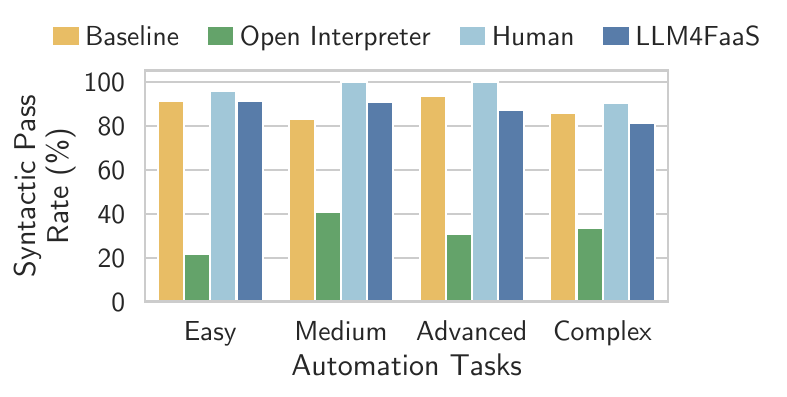}
    }
    \hfill
    \subfloat[Semantic pass rate\label{fig:eva:semantic}]{
        \includegraphics[width=0.49\linewidth]{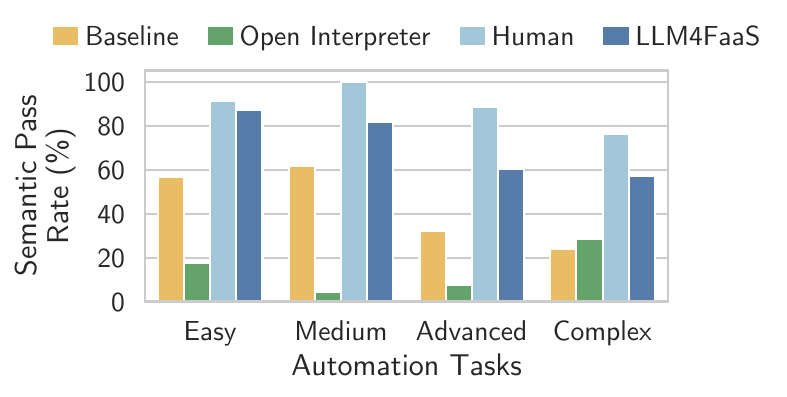}
        }
    \caption{
        Syntactic and semantic pass rates across four task complexities.
        We compare \sysname{} with (i)~a baseline without FaaS integration, (ii)~Open Interpreter, and (iii)~human developer.
        GPT-4o is used for all LLM-based experiments.
    }
    \label{fig:eva-llm4faas-baseline}
\end{figure*}

\subsection{Proof-of-Concept Implementation}
\label{sec:eva:poc}
To show the feasibility of the proposed system design, we implemented a proof-of-concept prototype of \sysname{} that we make available as open-source software.\footnote{\url{https://github.com/Mhwwww/LLM4FaaS}}
We also provide details on the system prompt in the prototype repository.
The prototype is implemented in Python that realizes the bridge system described in \S\ref{sec:architecture}.
We integrate OpenAI's GPT-4o model~\cite{hurst2024gpt} via the API access, setting the temperature to 0.7 and the maximum token limit to 1,500.
Our prototype uses \emph{tinyFaaS}~\cite{paper_pfandzelter2020_tinyfaas}, an open-source FaaS platform, to deploy and manage functions.
In our implementation, we use the tinyFaaS Python runtime and accordingly configure the LLM to generate Python code.
While our prototype integrates with GPT-4o, the design of \sysname{} is model-agnostic and can interface with other LLM APIs.

\subsection{Study Design}
\label{sec:eva:study-design}
Our evaluation uses a dataset of natural language application descriptions collected from real users.
This dataset serves as the basis for a series of experiments with our \sysname{} prototype.

\subsubsection{Dataset}
We collect the dataset through a questionnaire completed by 26 users without prior programming experience. 
In the questionnaire, we ask users to describe in natural language how they would instruct a smart home system to automate 4 tasks with increasing complexity.
We denote these tasks as \emph{easy}, \emph{medium}, \emph{advanced}, and \emph{complex} in the remainder of this paper.
We provide a detailed description on the design of our questionnaire and data collection in the dataset repository.\footnote{\url{https://github.com/Mhwwww/LLM4FaaS-dataset}}
All user answers are originally written in Chinese. 
To preserve the intended meaning and nuance of the user responses, we do not translate the answers for evaluation, yet we are aware that this can impact our results (see our discussion in \S\ref{sec:eva:discussion}).
In our system prompt, we combine these answers with API descriptions for our fictional smart home environment.
To support accessibility, we include English translation alongside the original Chinese in the dataset repository.

\subsubsection{Experiments}
We use our dataset of natural language descriptions to evaluate the efficacy of \sysname{}.
\begin{itemize}
    \item Syntactic pass: the application runs without errors or exceptions.
    \item Semantic pass: the application passes all functional tests based on the user intent.
\end{itemize}

Note that (i)~syntactic pass is a prerequisite for semantic pass and (ii)~a result may also be semantically incorrect if the user fails to understand or articulate the task functionality.
This is by design: Although this would not be the fault of the LLM, we do consider the \sysname{} approach to fail if the user is unable to build the application they desire.
We further discuss this in \S\ref{sec:eva:discussion}.

To the best of our knowledge, there is not a directly comparable end-to-end LLM-based no-code platform specializing for non-technical users in application customization.
To this end, we construct a three-dimensional evaluation to showcase the effectiveness of \sysname{} in bridging the gap between natural language programming and deployable customized solutions.

\begin{enumerate}
    \item \emph{Baseline}:
         We run \sysname{} without integrating FaaS, where the LLM generates \emph{applications} and CLI instructions to isolate and assess the impact of FaaS abstraction.

    \item \emph{Existing LLM-based platform}:
          We evaluate \emph{Open Interpreter}~\cite{open_interpreter}, an open-source LLM-based code generation and execution platform, to compare its operational simplifications with the infrastructure management provided by FaaS platform in \sysname{}. 
          
    \item \emph{Human developer}: 
        We ask a human developer to manually implement the applications from the user descriptions, to assess how vague expressions in the descriptions affect the correctness of LLM-generated results.  
\end{enumerate}

The human developer uses the same prompt as \sysname{}, i.e., developing only the required function logic.
For baseline and the Open Interpreter experiment, we modify the prompt to instruct the LLM generate the entire application, including the boilerplate code, and command line instructions to run the application.
While this is unrealistic, as it requires manual work by us to build and run the generated applications, it allows us to evaluate to what extent FaaS can actually reduce operational overhead and improve the practicality of LLM-generated results.

Finally, we also
(1)~repeat a subset of \sysname{} experiments to verify result stability,
(2)~assess code quality using Pylint and Radon,
(3)~evaluate \sysname{} performance in NodeJS to showcase the impact of the programming language, and 
(4)~measure the end-to-end latency across four different experiment setups,
providing a comprehensive view of \sysname{} performance.
We use OpenAI's GPT-4o model for all experiments involving LLM-based code generation. 

\subsection{Results}
\label{sec:eva:results}
We show the syntactic and semantic pass rate of \sysname{} alongside three comparative experiments in Figure~\ref{fig:eva-llm4faas-baseline}.
As both metrics quantify proportions of cases meeting the respective pass criteria, we omit error bars from the figures.

Across both metrics, the overall performance of the four experiments follows a descending order, i.e., human developer, \sysname{}, baseline, and Open Interpreter. 
The average syntactic pass rate for human developers, \sysname{} and the baseline are 96.53\%, 87.55\%, and 88.42\%, respectively, while Open Interpreter achieves 31.69\%.
Although the baseline achieves a marginally higher syntactic pass rate than \sysname{}, \sysname{} significantly outperforms the baseline in semantic, indicating its superior ability to generate functionally correct code.
Moreover, \sysname{} performs slightly better on easy and medium tasks, i.e., both above 90\%, and the baseline performs better on advanced and complex tasks.
The human developer achieves over 90\% syntactic pass rate across all tasks, reaching 100\% for the medium and advanced tasks.

For semantic pass, both human developer and \sysname{} maintain high average rates of 88.99\% and 71.47\%, respectively.
In contrast, the baseline drops to 43.48\%, and Open Interpreter further to 14.55\%.
For the easy and medium tasks, \sysname{} shows an over 80\% semantic pass rate, i.e., 86.96\% and 81.54\%, respectively, while it declines to around 60\% for advanced and complex tasks, i.e., 60.26\% and 57.14\%, respectively.
The human developer experiment has a 76.19\% semantic pass rate for the complex task, an around 90\% semantic pass rate for easy and advanced tasks, and reaches 100\% for medium task.

The high average semantic pass rate indicates that human developers generally understand user requirements, while the less-than-100\% results suggest that even human developer occasionally misunderstand or overlook certain aspects.
The semantic performance gap between human developer and \sysname{} suggests that failures stem from both misinterpreting the requirements and limitation in code generation.

\subsubsection{Syntactical Failure Reasons}
\label{sec:eva:results:syntactic-error}
\begin{table*}[h]
    \centering
    \begin{tabular}{lcccc}
        \toprule
        Experiment      & Pylint Score (avg ± std) & CC (avg ± std)  & MI (avg ± std)    & HALS Effort (median) \\
        \midrule
        Baseline         & $3.18 \pm 2.34$             & $8.51 \pm 6.99$             & $75.13 \pm 7.67$              & \textbf{75.31}                \\
        Human Developer  & $\bm{8.30 \pm 1.38}$  & $\bm{7.06 \pm 4.66}$             & $55.63 \pm 15.00$             & 272.32              \\
        \sysname{}       & $6.13 \pm 1.85$             & $9.43 \pm 7.42$             & $\bm{78.34 \pm 11.85}$             & 84.26                \\
        \textit{Open Interpreter} & $\mathit{7.25 \pm 2.06}$    & $\mathit{6.62 \pm 6.58}$  & $\mathit{79.73 \pm 12.50}$  & $\mathit{24.73}$              \\
        \bottomrule
    \end{tabular}
    \caption{
        Comparison of Code Quality Metrics across Experiments: 
        The table reports the average (avg) and standard deviation (std) of Pylint scores, Cyclomatic Complexity (CC), Maintainability Index (MI), and the median HALS Effort.
        Open Interpreter results are included for reference only.
        }
    \label{tab:eva:code_quality}
\end{table*}

\begin{figure*}
    \centering
    \subfloat[Syntactic pass rate\label{fig:eva:nodejs_syntactic}]{
        \includegraphics[width=0.48\linewidth]{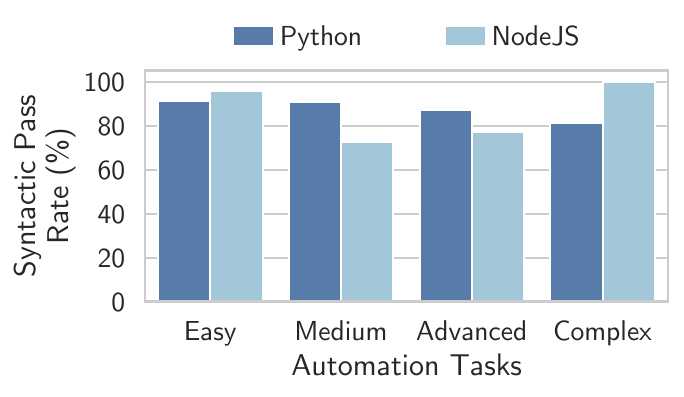}
        }
    \hfill
    \subfloat[Semantic pass rate\label{fig:eva:nodejs_semantic}]{
        \includegraphics[width=0.48\linewidth]{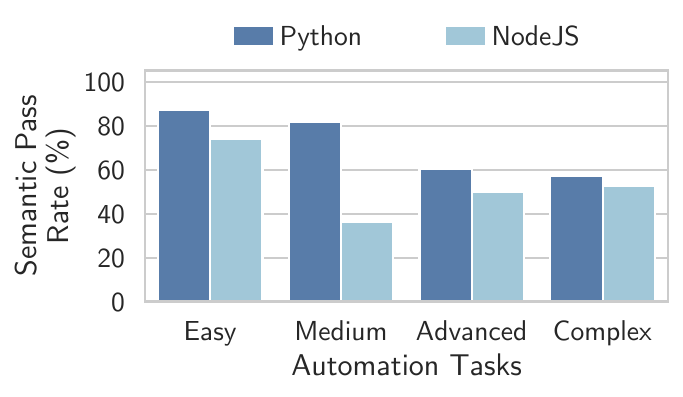}
    }
    
    \caption{
        Comparison of \sysname{} syntactic and semantic pass rates between NodeJS and Python.
        Syntactic pass rates remain consistently high in both case, while the semantic pass rate decreases for all task complexities under the NodeJS setup.
        }
    \label{fig:eva:nodejs}
\end{figure*}

\sysname{} and the baseline exhibit similar rates of syntactical failures, primarily stemming from import errors, improper data handling, and missing code. 
Import errors dominate, representing 81.82\% of \sysname{} errors and 57.14\% of baseline errors. 
For \sysname{} errors, 13.64\% are due to improper data handling, and 4.55\% are due to missing code.
In baseline, 28.57\% of errors are from missing code and 14.29\% due to improper data handling.
For Open Interpreter, 51.93\% of responses fail to produce code, resulting in a syntactical failure.
Although Open Interpreter is intended to generate and execute code locally, actual local file generation occurs only in 22.12\% of cases, while the remaining code-available responses require manual extraction.
In addition to the errors observed in other experiments, 33.65\% of responses report API rate limit errors, even though all experiments use the same model configuration and OpenAI account.
Notably, 40\% of the rate-limit-error cases succeed in generating local files, indicating that the error may arise during the code validation phase, whether in syntactic parsing or semantic checking.
For the human developer experiment, we do expect that it would have a high syntactic pass rate, as developer tests the code before submitting.
The few syntactical failure cases occur because the vague requirements are not sufficient for application development, resulting in either no code being generated or only skeleton code being provided.

\subsubsection{Quality of Generated Code}
\label{sec:eva:results:quality}

Syntactic and semantic passes can be achieved with varying levels of code quality.
Therefore, we further evaluate the code quality of the generated results in terms of complexity and maintainability.
We use Pylint\footnote{\url{https://www.pylint.org/}} and Radon\footnote{\url{https://pypi.org/project/radon/}} for assessment, getting an overall score and evaluating key metrics, i.e., cyclomatic complexity (CC), maintainability index (MI), and a module-level Halstead complexity in effort (Hals). 
We report the average and standard deviation for most metrics, while Hals is represented by its median value due to its skewed distribution.
Our comparison includes three experimental results, i.e., \sysname{}, baseline, and the human developer experiment, and presents the results in Table~\ref{tab:eva:code_quality}.
We include the Open Interpreter result for completeness but exclude it for comparison because of (i) its low pass rate in both syntactic and semantic aspects, and (ii) its limited code generation capabilities, i.e., only 34.62\% the responses yield valid Radon scores.


The human developer experiment achieves the highest Pylint score (8.30), indicating high code quality with fewer stylistic and structural issues. 
\sysname{} ranks second (6.13), while the baseline experiment scores lower (3.18), showing that LLMs can generate syntactically sound code, but further refinements is necessary to reach a human-like code quality.
This result also suggest that by leveraging FaaS, \sysname{} enables LLM to produce code that is more structured and maintainable.

The human developer results exhibit the lowest cyclomatic complexity while \sysname{} has the highest followed by the baseline.
This suggests that while \sysname{} produces functionally correct code, it tends to introduce more branching and decision points, increasing structural complexity.
\sysname{} achieves the highest maintainability score, followed by the baseline and human-written code. 
The relatively lower MI of human-written code is likely due to its complexity and use of idiomatic programming patterns that are harder to quantify in maintainability formulas.
Human developed code has the highest Halstead effort, whereas \sysname{} and the baseline, i.e., the LLM-generated results, require significantly lower values. 
This indicates that human-written code, while typically of higher quality, can be more complex and difficult to modify due to nuanced design decisions and domain-specific optimizations.

The code quality results suggest that \sysname{} effectively balances automation with code quality, producing maintainable code while reducing cognitive overhead compared to human-written solutions.
However, refinements in structure and complexity management can further enhance its usability in real-world scenarios.

\subsubsection{Programming Language Experiment}
\label{sec:eva:results:nodejs}

LLMs can perform differently across programming languages. 
To evaluate cross-language performance, we 
(i)~reimplement the fictional smart home application in NodeJS, 
(ii)~adapt the prompt to generate NodeJS functions, and 
(iii)~replace API descriptions with the NodeJS equivalents.

The results in Figure~\ref{fig:eva:nodejs} show that syntactic pass rate of \sysname{} remains high across programming languages despite fluctuating with task complexity, averaging 87.55\% for Python and 86.33\% for NodeJS.
In contrast, the semantic pass rate declines notably in NodeJS, i.e., 53.16\% on average, compared to 71.47\% using Python, suggesting that language-specific factors affect the LLM's ability to generate semantically correct code.

Notably, both the syntactic and semantic pass rate of NodeJS results deviate from task complexity, i.e., easy and complex tasks outperform the medium and advanced ones, with the complex task even achieving a 100\% syntactic pass rate. 
A potential reason lies in how the complexity is introduced and its interaction with the language-specific LLM capabilities.
In medium and advanced tasks, complexity stems from three subtasks, where advanced task has a higher internal complexity, whereas the complex task involves reasoning through a multifaceted scenario without subtasks.
The presence of subtasks may hinder the LLM ability to generate syntactically and semantically correct NodeJS code.

\subsubsection{Latency Experiments}
\label{sec:eva:result:latency}

\begin{figure}
    \centering
    \includegraphics[width=0.49\textwidth]{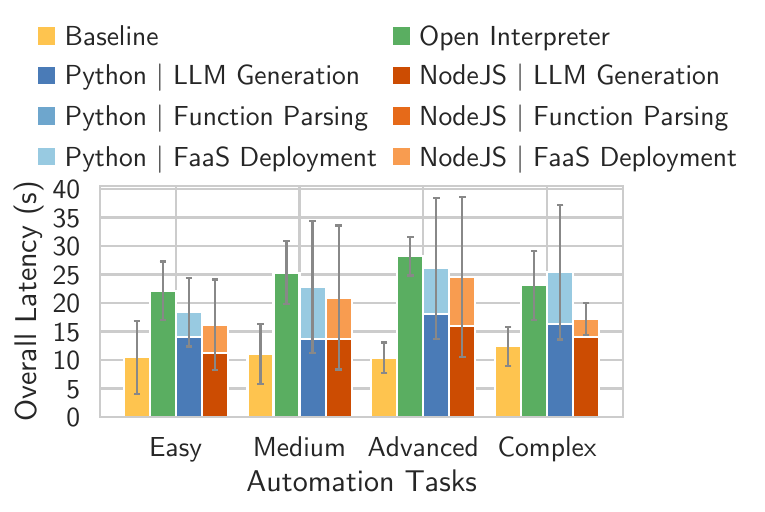}
    \caption{
        Average end-to-end latency of all experiments, decomposed into LLM generation, FaaS preparation, and deployment. 
        Error bars indicate standard deviation.
    }
    \label{fig:eva:latency-all}
\end{figure}

We evaluate the end-to-end latency of \sysname{} to assess its responsiveness as a complete system, from request initiation to successfully FaaS deployment.
For comparison, we also measure the latency for the baseline and Open Interpreter, defined as the time from sending the structured prompt to the LLM until the process completes, either successfully or with an exception.
We exclude the human developer experiment due to its non-automated nature  making direct latency measurement impractical.
We show the results in Figure~\ref{fig:eva:latency-all}.

\sysname{} latency comprises three stages, i.e., LLM generation, function preparation, and FaaS deployment.
When generating code in Python, the average end-to-end latency of \sysname{} among four tasks increases with task complexity, i.e., 18.40, 22.85, 26.08, and 25.37 s, respectively.
LLM generation dominates the end-to-end latency, averaging 13.87 s for simpler tasks, i.e., easy and medium tasks, and 17.19 s for more complex ones, while deployment ranges from 4.31 s to 9.20 s.
Function parsing time remains nearly constant and negligible, averaging 8.51 \emph{ms}.
When generating code in NodeJS, \sysname{} shows an average overall latency of 19.75 s, following the same distribution as in Python, i.e., LLM generation dominates, deployment follows, and packaging remains negligible at milliseconds.
Furthermore, in NodeJS, latency pattern mirrors semantic pass rates, i.e., easy and complex tasks outperform medium and advanced ones.
Although NodeJS reduces both generation and deployment by around 1 s compared to Python, this likely reflects less refined code. 
As baseline requires manual execution and deployment of the generated code, we compare it with the LLM generation duration of \sysname{} in the Python implementation.
The baseline shows an average latency of 11.03 s compared to 15.53 s for \sysname{}.
Given the lower semantic pass rate of baseline, we suggest that the shorter latency results from less refined code generation, which typically requires less processing time.
Open Interpreter exhibits the longest latency for all tasks except for the complex one, and 39.13\% of responses reached the 30-second timeout, suggesting that actual latency can be longer with a higher timeout window.

\subsubsection{Repeat Experiment}
\label{sec:eva:results:repeat}

\begin{figure*}
    \centering
    \subfloat[Syntactic repeat pass rate\label{fig:eva:repeat_syntactic}]{
        \includegraphics[width=0.48\linewidth]{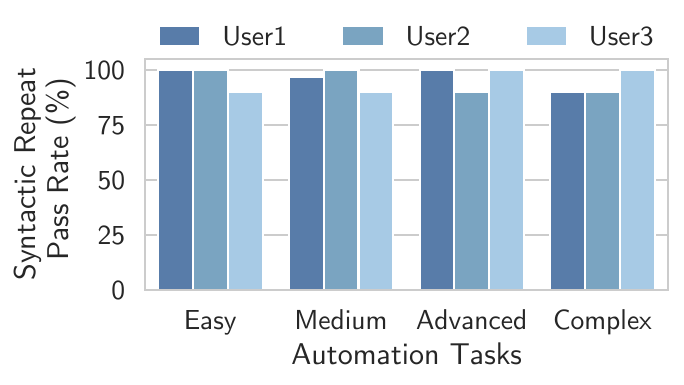}
    }
    \hfill
    \subfloat[Semantic repeat pass rate\label{fig:eva:repeat_semantic}]{\includegraphics[width=0.48\linewidth]{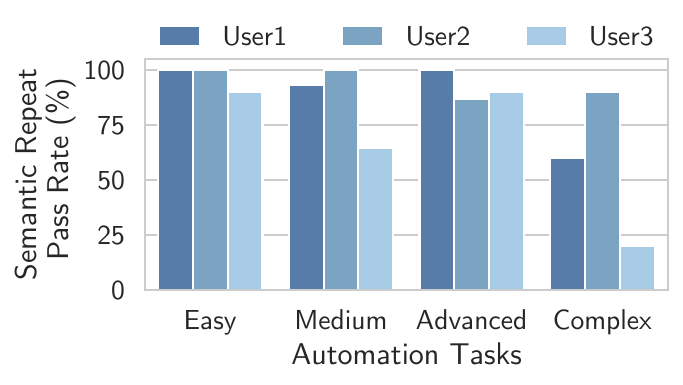}
    }
    \caption{
        Pass rates of ten repetitions of \sysname{} with three user answers.
        Syntactic pass remains stable between repeat LLM invocations (nine or ten of ten correct).
        For the complex task, repeating an identical invocation can lead to different results.
    }
    \label{fig:eva:repeat}
\end{figure*}

We use three random user answers to evaluate performance variation in \sysname{} with Python implementation.
Specifically, we aim to quantify to what extent randomness in LLM code synthesis affects the performance of \sysname{}.
To that end, we invoke \sysname{} ten times with identical natural language descriptions, recording syntactic and semantic pass outcomes for each repetition.

The results in Figure~\ref{fig:eva:repeat} show that syntactic pass rate remains stable across repetitions, with at most one failure out of ten. 
The semantic pass rate, however, exhibits greater variability for more complex tasks.
This shows how randomness in LLM responses, even with identical requirements, can impact the performance of \sysname{}.
Note that for semantic correctness, especially, it is not feasible to simply repeat an invocation until success, as there exists no way to automatically confirm correctness (unlike with syntactic correctness, where, e.g., import errors can be detected).

\subsection{Discussion}
\label{sec:eva:discussion}
Our prototype and experimental results demonstrate that the \sysname{} is feasible, especially for tasks of low complexity. 
Building on these, we next discuss its implications and limitations.

\subsubsection{Impact of FaaS}
Results from the baseline and Open Interpreter experiments indicate that leveraging FaaS in \sysname{} effectively preserves the syntactic correctness of LLM-generated code.
Moreover, the reduced complexity from FaaS programming model allows the LLM to focus on understanding and implementing user intentions rather than boilerplate code for operational concerns, leading to a significantly higher semantic pass rate.

\subsubsection{Task Complexity}
Our results show that task complexity impacts the semantic pass rate of \sysname{}.
We consider two possible causes:
First, the increased task complexity presents challenges for LLMs, posing more opportunity for failure.
Second, increased complexity can also be challenging for users, particularly those with less experience, who must understand and articulate more complex requirements.
The human developer experiment results further support this view.
Future work on prompting strategies could help LLM4FaaS better handle complex, multi-intent tasks.

\subsubsection{Model Selection}
We choose GPT-4o as the primary LLM for \sysname{} due to its advanced capabilities and strong performance in non-English languages~\cite{hurst2024gpt}.
\sysname{} exhibits practical performance with GPT-4o, even when handling more complex tasks.
Nevertheless, advances in LLMs will likely improve the performance of \sysname{}.
In future work, we plan to further evaluate our approach with different LLMs, including models specifically trained for software development.

\subsubsection{Size of the User Group}
In this paper, our evaluation is based on answers from 26 non-technical users.
While it would have been desirable to work with a larger user group for our experiments, this was not feasible due to the resulting effort in acquiring additional users and handling their answers which both required significant manual effort.
Nevertheless, we believe that our experiments reliably demonstrate the effectiveness of \sysname{} as this depends less on precise numbers but more on a general value range -- and for this our user group was large enough.

\subsubsection{User Description Language}
The natural language application descriptions collected from our users are all in Chinese, which may negatively affect the performance of the LLM.
While we cannot expect users to learn English to use a no-code development platform, in the same way that we cannot expect them to learn a programming language, we should be aware of the impact of input languages.
In future work, we plan to further investigate this impact by (i)~exploring language-specific LLMs, e.g., those trained on mostly Chinese-language texts, and (ii)~evaluating the feasibility of adding a separate translation step, despite the potential loss of nuance.

\subsubsection{Programming Language}
Similar to user description language, the choice of programming language in evaluation can also influence the performance of \sysname{} prototype.
We choose Python as primarily experimental setup due to its widespread adoption, preference by LLMs, and potential strong performance~\cite{twist2025llms}.
To investigate the impact of programming language on \sysname{}, we reimplement the same fictional smart home application in NodeJS, another widely adopted language, and compare its performance with the Python version.
Experimental results indicate that Python is better suited to \sysname{}, achieving a higher semantic pass rate with relatively low latency.
We leave the exploration of additional language backends as future work.

\subsubsection{Feasibility of Feedback Loops}
In \sysname{}, we give LLMs only a single opportunity to generate function code, without involving verification process of either with users or runtime errors.
However, LLMs are known to perform well with feedback.
It may be equally possible to provide feedback to the LLM on generated code, both from the FaaS platform for syntax errors~\cite{wang2025exploring}, and from the user for semantic errors, e.g., clarifying application logic.
However, given the latency concern and the rate limit errors observed in the Open Interpreter experiment, the feedback loops may inherently introduce throughput bottlenecks due to rate limit policies of LLM services. 
This highlights the need to balance the trade-off between code accuracy and service availability when designing interactive generation pipelines.

\subsubsection{Practical Deployment Constraints}
\sysname{} inherits operational constraints from its underlying FaaS platform that affect deployment flexibility. 
The stateless nature of FaaS limits data persistence across invocations, posing challenges for repetitive or context-dependent tasks.
Future work could enable \sysname{} to infer when implicit state retention is required from user intent and offer configurable persistence options within the deployment pipeline.
Also, cold-start delays caused by container initialization and LLM inference slow down the first invocation, particularly for newly deployed functions. 
Future work may explore caching or pre-warming strategies to mitigate these overheads, improve responsiveness.

%% file: sections/6_conclusion.tex
\section{Conclusion}
\label{sec:conclusion}

LLMs are powerful tools for generating code from natural language descriptions, but their adoption by non-technical users is hindered by the complexity of application deployment and operation.
With \sysname{}, we have proposed leveraging the high levels of abstraction offered by the FaaS paradigm to handle operation and code execution for non-technical users.
Furthermore, we proposed to leverage the reduced complexity of the FaaS programming model to improve correctness of LLM-generated applications.
 
We demonstrate the feasibility of \sysname{} with a proof-of-concept prototype and a new dataset of real user application descriptions which we make available as open source/data.
Experimental results show that \sysname{} achieves high reliability, efficiency and availability while maintaining competitive code quality, making it a viable solution for non-technical users to develop and run tailored applications without requiring any technical expertise.